\documentclass[a4paper,superscriptaddress]{revtex4}

\pdfoutput=1
\usepackage{graphicx}

\usepackage{enumitem}
\setdescription{leftmargin=0pt}

\usepackage{soul}

\usepackage[english]{babel}
\usepackage[T1]{fontenc}
\usepackage{soul}

\usepackage{amssymb}
\usepackage{siunitx}
\usepackage[usenames,dvipsnames]{xcolor}

\newcommand{\ds}{\ensuremath{\displaystyle}}
\newcommand{\etahot}{\ensuremath{\eta_{\,\rm hot}}}

\newcommand{\affloa}{\affiliation{Laboratoire d'Optique Appliqu\'ee, ENSTA-ParisTech, Ecole
Polytechnique-ParisTech, CNRS UMR 7639, 828 Boulevard des Marechaux, 91762 Palaiseau CEDEX, France}
}
\newcommand{\affgolp}{\affiliation{GoLP/Instituto de Plasmas e Fusão Nuclear, Instituto Superior
Técnico, Universidade de Lisboa, Lisbon, Portugal} }

\begin{document}

%% title, 15w
%\title{\flushleft \Large Magnetic fields in plasma during high-intensity laser pulse propagation}
\title{Persistence of magnetic field driven by relativistic electrons in a plasma}

\author{A. Flacco}\email{alessandro.flacco@polytechnique.edu}\affloa{}

\author{J. Vieira}\affgolp{}
\author{A. Lifschitz}
\author{F. Sylla}
\author{S. Kahaly}
\author{M. Veltcheva}\affloa{}
\author{L. O. Silva}\affgolp{}
\author{V. Malka}\affloa{}

%% abstract 200w, full refs
\begin{abstract}
The onset and evolution of magnetic fields in laboratory and astrophysical plasmas is determined by
several mechanisms~\cite{bib:zweibel_rpp_2008}, including instabilities
\cite{bib:medvedev_apj_1999,bib:silva_apjl_2003}, dynamo
effects~\cite{bib:brandenburg_ssr_2012,bib:russel_apj_1992} and ultra-high energy particle flows
through gas, plasma and interstellar-media~\cite{bib:miniati_apj_2011,bib:gregori_nature_2012}.
These processes are relevant over a wide range of conditions, from cosmic ray acceleration and gamma
ray bursts to nuclear fusion in stars. The disparate temporal and spatial scales where each operates
can be reconciled by scaling parameters that enable to recreate astrophysical conditions in the
laboratory.  Here we unveil a new mechanism by which the flow of ultra-energetic particles can
strongly magnetize the boundary between the plasma and the non-ionized gas to magnetic fields up to
10-100 Tesla (micro Tesla in astrophysical conditions). The physics is observed from the first
time-resolved large scale magnetic field measurements obtained in a laser wakefield accelerator.
Particle-in-cell simulations capturing the global plasma and field dynamics over the full plasma
length confirm the experimental measurements. These results open new paths for the exploration and
modelling of ultra high energy particle driven magnetic field generation in the laboratory.

%% --- 177 words ---

\end{abstract}

\maketitle

\setlength{\unitlength}{0.1\linewidth}
\begin{figure*}[!ht]
\begin{center}
\begin{picture}(10,3.5)
	\put(1.4,0.5){\includegraphics[height=4cm]{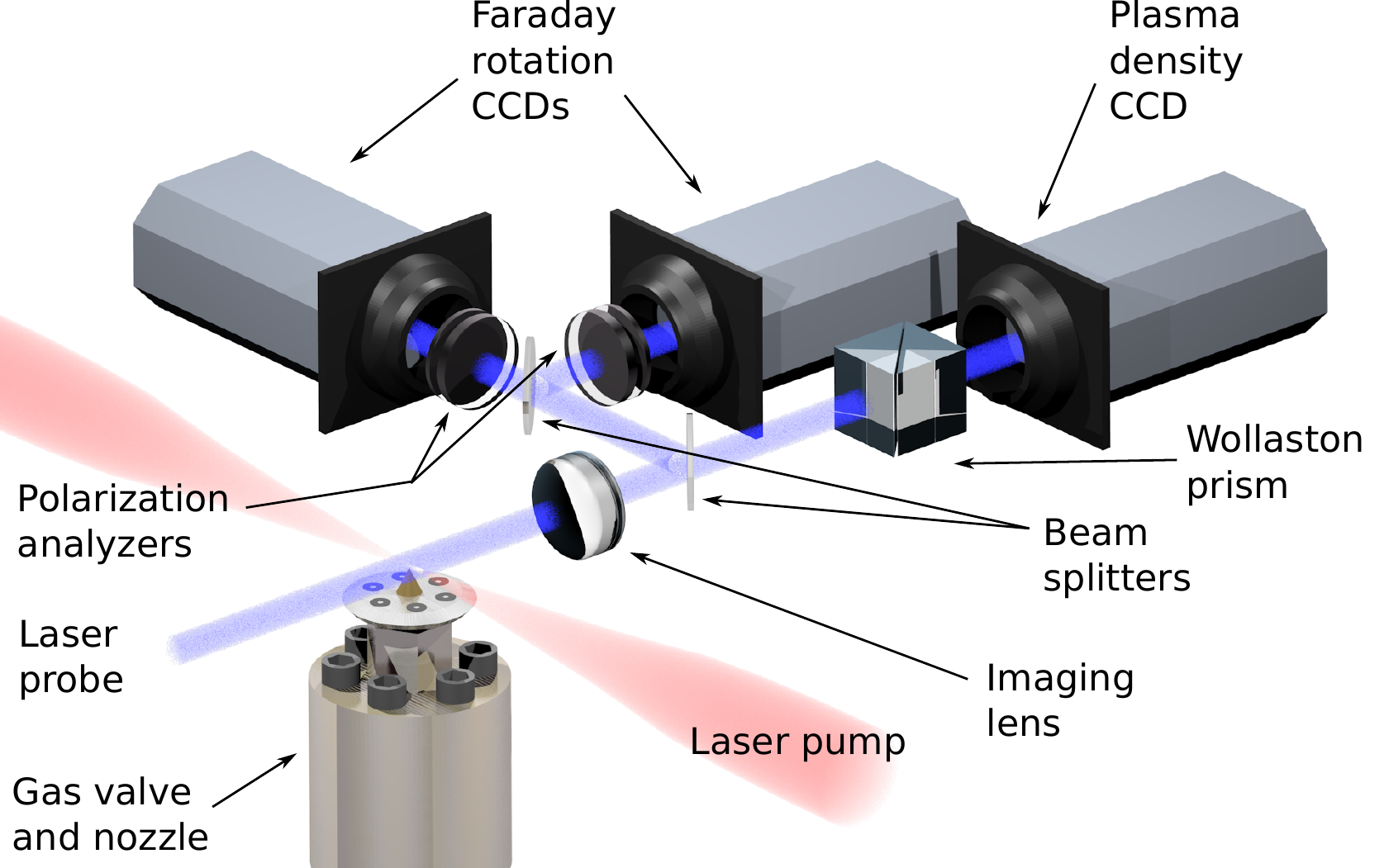}\hspace{1cm}}
	\put(1,2.8){(a)}
	\put(5.7,0){\includegraphics[height=5.7cm]{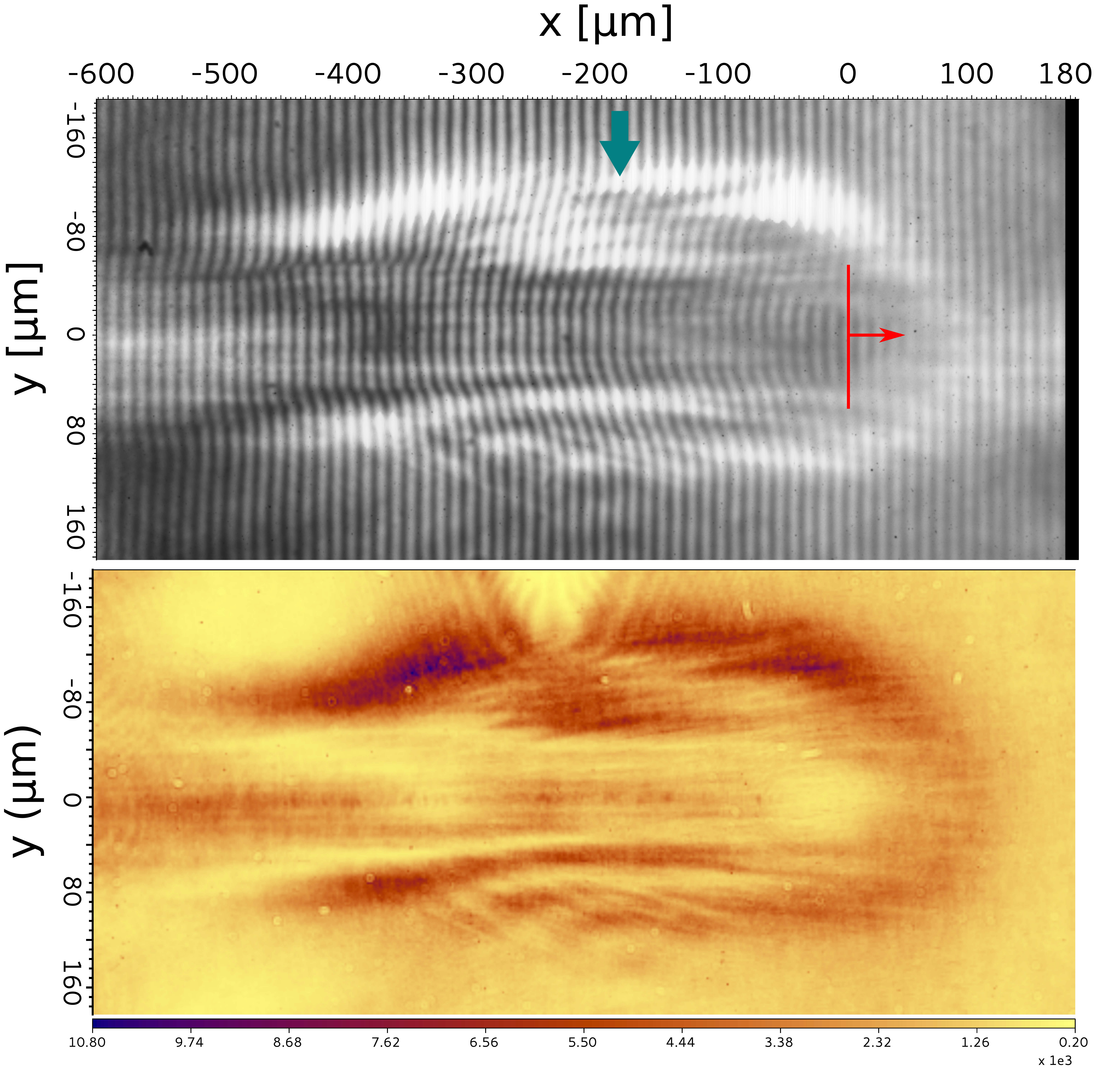}}
	\put(7.6,2.8){\tiny\color{white}\bf Gas flow}
	\put(8.3,2.4){\tiny Laser}
	\put(9.1,2.8){(b)}
	\put(9.1,1.3){(c)}
%\put(10,3.5){x}
\end{picture}
\end{center}
\caption{ {\bf a}: Scheme of the experimental setup, showing the laser pump beam (red) focused into
the gas jet and the probe beam (blue) split multiple times for simultaneous measurement of electron
density and Faraday rotation in the plasma.  The probe beam propagates perpendicularly to the pump
laser path, integrating the polar magnetic component and the density in the plasma. {\bf b,c}:
Images of the plasma while the laser propagates at $t=\SI{1.9}{\ps}$. {\bf b}:  interference map
showing the integrated phase. {\bf c}: one of the two simultaneous images recording the polarization
rotation onto the intensity (false colors).}
\label{fig:scheme}
\end{figure*}
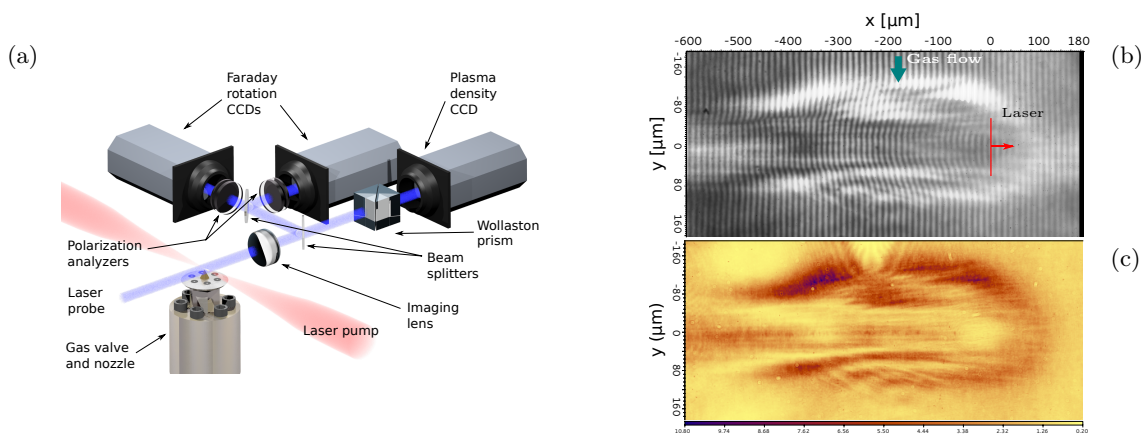

Strong cosmic magnetization requires ultra-high energy particle flows. These non-thermal particle
streams can be originated~\cite{bib:hillas_araa_1984} by statistical acceleration processes, such as
shock~\cite{bib:spitkovsky_apjl_2008,bib:martins_apjl_2009} and Fermi
acceleration~\cite{bib:fermi_pr_1949}, and by direct particle acceleration mechanisms in strong wave
fields, such as those found in pulsars~\cite{bib:gunn_prl_1969}. In addition to direct astronomical
observations, the physical processes occurring when these mechanisms take place may also be explored
in the laboratory. For instance, the conditions for the onset of statistical acceleration mechanisms
through collisionless shocks in the laboratory have been investigated theoretically~\cite{bib:fiuza_prl_2012}. Direct cosmic acceleration can also be explored in the laboratory through
laser driven plasma wakefields~\cite{bib:rosensweig_pra_1988,bib:chen_prl_2002,bib:yin_prl_2009}.
Here we show that the non-thermal particle flows produced in a laser wakefield accelerator
(LWFA)~\cite{bib:malka_science_2002, bib:modena_nature_1995} can strongly magnetize the plasma and
the plasma-neutral gas boundary. This observation is also the first time the strong magnetization
occurring at the flow of energetic particles from ionised to non-ionised interstellar material can
be reproduced in the laboratory.

A laser wakefield accelerator (LWFA) uses short and intense laser pulses to drive large amplitude
plasma waves.  In the LWFA scheme, the laser ponderomotive force excites ultra-relativistic, large
amplitude plasma waves where electrons can be trapped and accelerated to high energies. In our
experiment we used a \SI{30}{\fs} laser pulse focused to a transverse spot-size of
\SI{8}{\micro\meter} (FWHM) for a peak intensity of I$_{0} = \SI{3e19}{\watt\per\centi\meter^{2}}$.
The plasma is created in an Helium gas jet at near critical density, $n_A =
\SI{3.5e19}{atoms\per\cm^{3}}$.  Since the laser power is above the critical power for
self-focusing, it excites strongly non-linear plasma waves above the wave-breaking threshold.
Wave-breaking leads to non-thermal particle flows with relativistic energies.

Energetic particles produced during wavebreaking propagate mainly in the forward (laser) direction.
However, a fraction of these particles also propagates radially to regions of undisturbed plasma,
eventually reaching the plasma-neutral gas boundary. As they propagate through the ionised medium,
return currents are set up to balance the hot electron flow, preventing effective magnetic field
generation.  However, as the hot electrons cross the plasma-neutral gas boundary, strong magnetic
fields are induced by the resulting current imbalance. We demonstrate that this mechanism can
produce strong magnetic fields scaling with $\ds 32\etahot\left(n_0
[10^{16}\si{\cm^{-3}}]\right)^{1/2}\si{\tesla}$ in the laboratory. In astrophysical conditions,
fields created by the same mechanism can reach amplitudes of $\ds \left(0.1-1\right) \etahot
\left(n_0[\si{cm^{-3}}]\right)^{1/2}\si{\micro\tesla}$, where $\etahot$ is the fraction of hot
electrons/energetic particles to the background plasma density $n_0$.

Our experiment provides the first time resolved measurements of the magnetic field spatial
distribution in a laser-plasma accelerator for the whole plasma volume and with a high temporal
resolution.  As the laser propagates through the gas target we observed strikingly complex magnetic
structures, with several inversions of the field orientation at the plasma core. These observations
are in excellent agreement with 3D one-to-one Particle-in-Cell (PIC) simulations in
OSIRIS~\cite{bib:fonseca_book,bib:fonseca_ppcf_2013} capturing the global plasma dynamics and
magnetic field evolution over the entire gas jet
each taking several hundreds of thousands of CPU-hours (see methods section).  

\begin{figure*}[!ht]
\begin{center}
\begin{minipage}{0.8\textwidth}
\includegraphics[height=4.8cm]{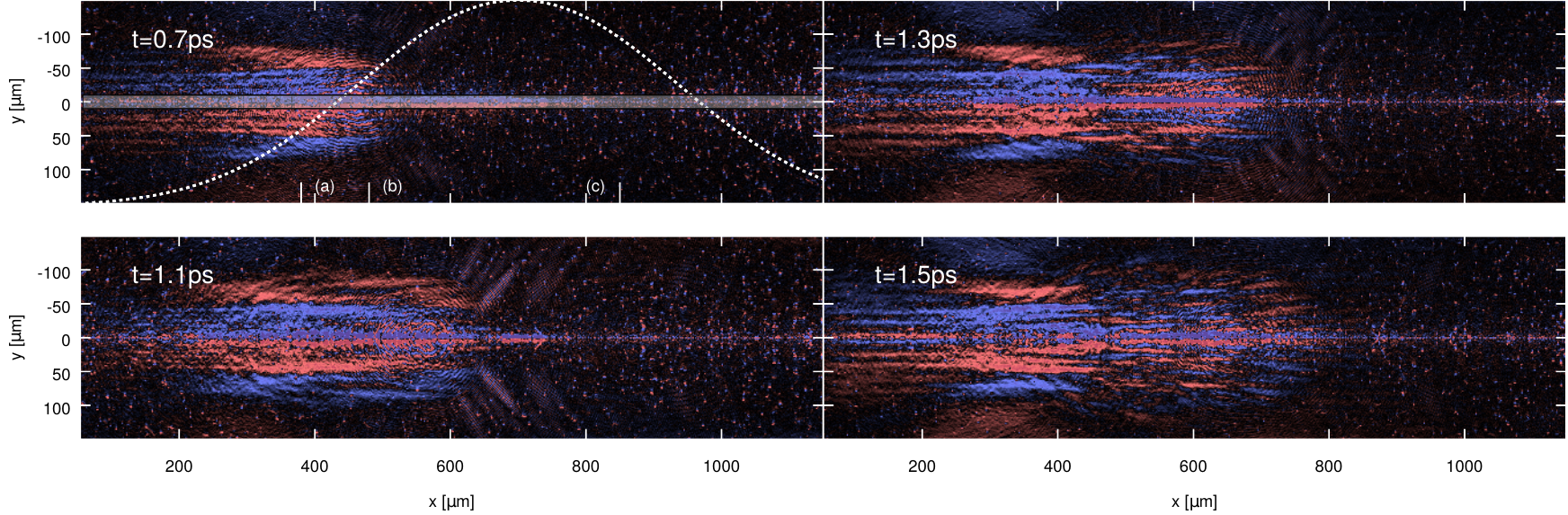}
\end{minipage}
\hspace{12mm}
\begin{minipage}{0.1\textwidth}\vspace{-7mm}
\includegraphics[height=4.1cm]{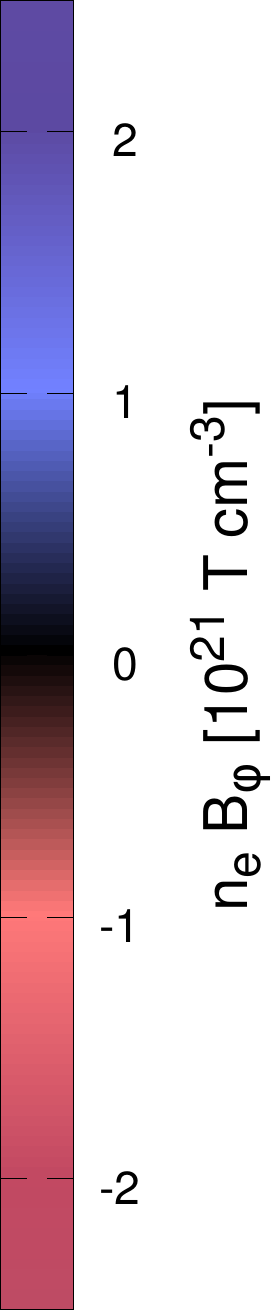}
\end{minipage}
\end{center}
\caption{Magnetization of the plasma at different times. The laser pulse enters the gas area at
$t=\SI{0.3}{ps}$ from the left hand side of the box. {\bf t=0.7ps}, the bell-shaped plot represents
the longitudinal electron density profile in the plasma, highest density being
$n_e=\SI{7e19}{cm^{-3}}$. The grayed region represents the beam waist projection.}
\label{fig:nebs}
\end{figure*}

The laser interaction with the gas jet is thoroughly scanned at high temporal and spatial
resolutions for electron density (via phase recording) and magnetic field mapping.  The plasma is
probed by a single $\SI{30}{fs}$ pulse, synchronized to the pump pulse and doubled in frequency, as
shown in Fig.\ref{fig:scheme},a. During its propagation in the plasma, polarization rotation and
absolute de-phasing are integrated and recorded on three separate CCD cameras.  The
three-dimensional density and magnetic maps are then reconstructed from the recorded information
(see Methods). This pump-probe experiment permits us to ``freeze'' the plasma state to a high
resolution snapshot lasting only for \SI{30}{\fs} and to follow its evolution.  The validity of our
diagnostic reposes on the assumption that ${\bf B}\cdot\nabla n_e=0$ in the plasma cylinder, which
is confirmed by simulations.

Snapshots of the spatial distribution of the  magnetic field  at selected times are shown in Figure
\ref{fig:nebs}.  Each image represents the symmetrized radial map of $\left( n_e {\bf
B}\right)_{\phi} \left( r, x \right)$, product between the polar component of the magnetic field and
the local electron density, as reconstructed from the probe polarization maps. The laser pulse
propagates from the left to the right.  A polar magnetic field is observed in the trail of the laser
pulse soon after its entrance in the gas jet ($t=\SI{0.7}{\ps}$). From the density measurement we
can infer a magnetic field magnitude reaching \SI{100}{\tesla}. This field is positive in the plasma
core ($r<\SI{50}{\um}$) and  changes in sign around the radial border of the plasma ($r \sim
\SI{50}{\um}$). As we will see from simulations below, the inversion  of the field direction at the
boundary  is a distinctive feature indicating an electron current passing trough the plasma/gas
boundary.  These electrons (ultra-high energetic particles in astrophysical scenarios) are
relativistic and, though accelerated from the wavebreaking, they are not trapped by the wakefield.
%%s

Figure~\ref{fig:nebs} shows  another striking feature, consisting in a strong magnetization of the
plasma core.  A large magnetic field is indeed expected close to the laser propagation axis, in the
wakefield region. This field is created by the very high longitudinal wakefield current, which is
not screened by the plasma.  At a distance corresponding to the limit of the laser beam waist (i.e.
a few \si{\um}'s), a weak magnetic field survives (the bow wave magnetic
field~\cite{bulanov_PRL_bow}).  Farther away from the axis, plasma return currents typically screen
hot electrons that expand radially from the wakefield (note that the main velocity component of
these electrons is along $x$), thus no significant field is expected to survive up to the boundary
whereabouts. The experimental results suggest that in the plasma core the screening of hot electron
current coming from the wakefield is not as efficient as expected.

The magnetic field structure remains alike until $t=\SI{1.1}{\ps}$: starting from $t=\SI{1.3}{\ps}$,
an island of reversed field starts building up close the gas jet density peak ($x=\SI{530}{\um}$),
relatively far from the laser waist. At $t=\SI{1.5}{\ps}$ (\ref{fig:nebs}), this island has expanded
longitudinally and radially, and a second one appears closer to the laser axis ($x=\SI{580}{\um}$),
at a larger radius. Simulations suggest that these islands are caused by the filamentation of the
wings of the laser pulse, i.e. the laser energy surrounding the central spot.  

\begin{figure*}[t]
\begin{center}
\includegraphics[width=0.7\textwidth]{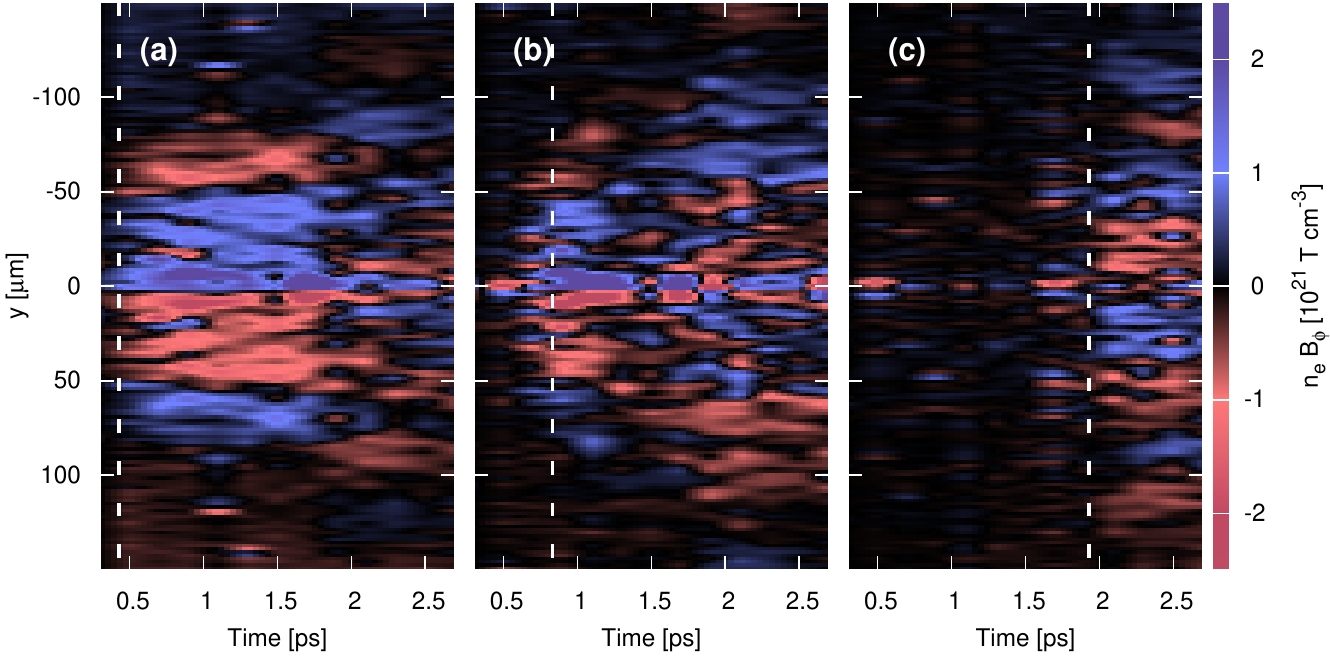}
\end{center}
\caption{Composed images showing the evolution in time of the magnetic field (temporal resolution:
\SI{200}{\fs}). The three panels correspond to {\bf a, b, c} markers in Fig.\ref{fig:nebs}. Dotted
lines indicate the laser arrival.} 
\label{fig:neb_evolution}
\end{figure*}

Experiments showed several inversions of the magnetic field orientation as illustrated in
Fig.~\ref{fig:neb_evolution} Several shots at different delays are used to compose a picture of the
local evolution in time of the magnetic field. The temporal resolution in this figure is of
\SI{200}{\femto\second}.  Upon its entrance in the gas (Fig.\ref{fig:neb_evolution}a) the laser
propagation is accompanied by the formation of a positive poloidal magnetic field in the plasma
core, which changes sign in the plasma boundary. The inner field component is consistent, in sign,
with a negative current propagating with the laser pulse and remains stable for approximatively
\SI{1.2}{\ps}.  Panel {\bf b} of figure \ref{fig:neb_evolution} shows the evolution in the region
where the first island appears, $x=\SI{480}{\um}$. At $t\sim\SI{1.1}{\ps}$ the field sign is
reversed in the plasma core, corresponding to the formation of the first island. This inversion
lasts up to $t \sim \SI{1.5}{ps}$.  When the laser exits from the gas profile
(Fig.\ref{fig:neb_evolution}c) a poloidal field is reformed, comparable to panel (a). At this time,
however, the plasma core magnetisation is opposite, in sign, to what was observed at the entrance.

Experimental findings are confirmed by 3D PIC simulations run with parameters closely matching
experimental laser and plasma conditions (see methods).  As it enters the gas, the laser ionizes the
gas up to a radius of \SI{100}{\um} away from the axis and excites weakly non-linear plasma waves.
Relativistic pulse self-focusing enhances the wakefield amplitude beyond the wavebreaking threshold
after \SI{400}{\um} of propagation.  When wavebreaking occurs, a fraction of the resulting hot
electrons expands radially through the plasma.

The laser driven plasma waves lead to complex longitudinal electron current structures in the
ionised volume. Figure \ref{fig:3dsims}-{\bf I} shows the current structures at the end of the
simulation, at $t=\SI{2.67}{\ps}$, where it is possible to distinguish between {\it backward
electron currents} (blue) and {\it forward electron currents},(red). As electrons cross the
plasma-neutral gas boundary an inner return current is set up, located at the boundary itself, at a
variable radius in the range $r \lesssim 60-100\,\si{\um}$.  These currents,
figure~\ref{fig:3dsims}-\textbf{I}, are at the origin of the large scale poloidal magnetic fields
observed in figure \ref{fig:3dsims}-\textbf{II}. 
Simulations also show that a small fraction of hot electrons at the plasma entrance
($x<\SI{500}{\um}$) flow away from the laser in the backward direction. These are indicated by the red structures involving a blue bulk ($r \gtrsim
\SI{60}{\um}$) in \mbox{Fig.  \ref{fig:3dsims}-I{\bf a}}. For $x>\SI{500}{\um}$ most of the hot
electrons are accelerated in the forward direction leading to the outer forward (blue) current
structures in \mbox{Fig.  \ref{fig:3dsims}-I {\bf b, c}}. 

Return currents electrons propagate forward at the leading edge of the gas profile
(Fig.~\ref{fig:3dsims}~Ia) and backward for the remainder of the gas jet length
(Fig.~\ref{fig:3dsims}~Ib-c).  Because hot electrons flow in opposite directions for
$x<\SI{500}{\um}$ (Fig.~\ref{fig:3dsims}~IIa) and for $x>\SI{500}{\um}$ (Fig.~\ref{fig:3dsims}~IIc),
the magnetic field sign changes in these regions.  The transition at $x\simeq \SI{500}{\um}$ is
shown in Fig.~\ref{fig:3dsims}~IIb, which illustrates the poloidal magnetic field corresponding to
the longitudinal currents shown in Fig.~\ref{fig:3dsims}~I. This transition to a more rich current
structure around $x\simeq\SI{500}{\um}$ is confirmed by experimental observation, e.g.
Fig.~\ref{fig:nebs} and Fig.~\ref{fig:neb_evolution}b.

The large scale magnetic field that surrounds the plasma around $|y|\simeq 60-100\,\si{\um}$ reaches
amplitudes that are in fair agreement with experimental measurements ($n_0 B_z\simeq
\SI{e21}{\tesla\per\cm^{3}}$). 
\setlength{\unitlength}{0.1\textwidth}
\begin{figure*}[!ht]
	\begin{center}
		\begin{picture}(10, 4.5)
			\put(0,0.15){\includegraphics[width=0.41\textwidth]{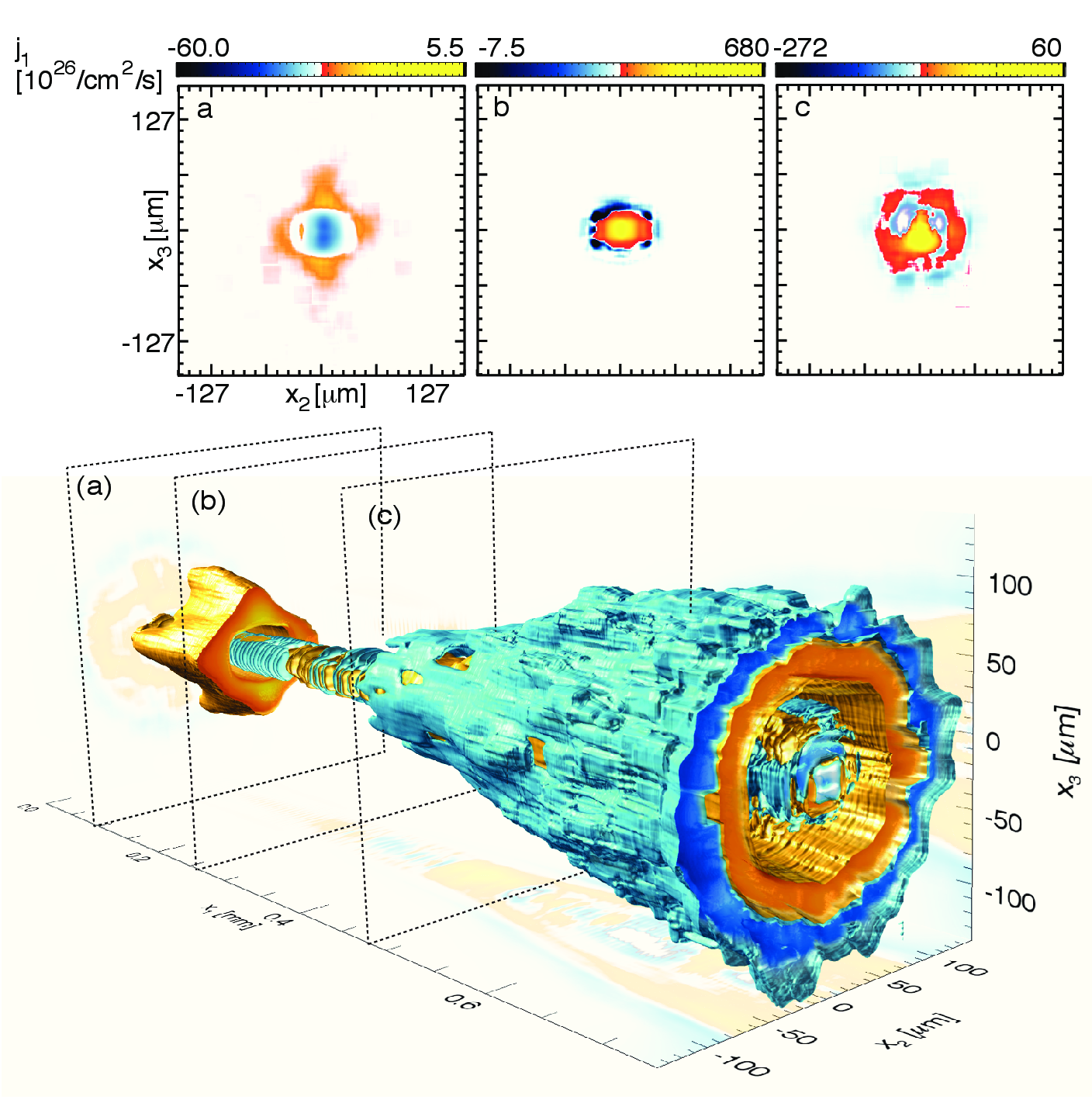}}
			\put(4.8,0){\includegraphics[width=0.455\textwidth]{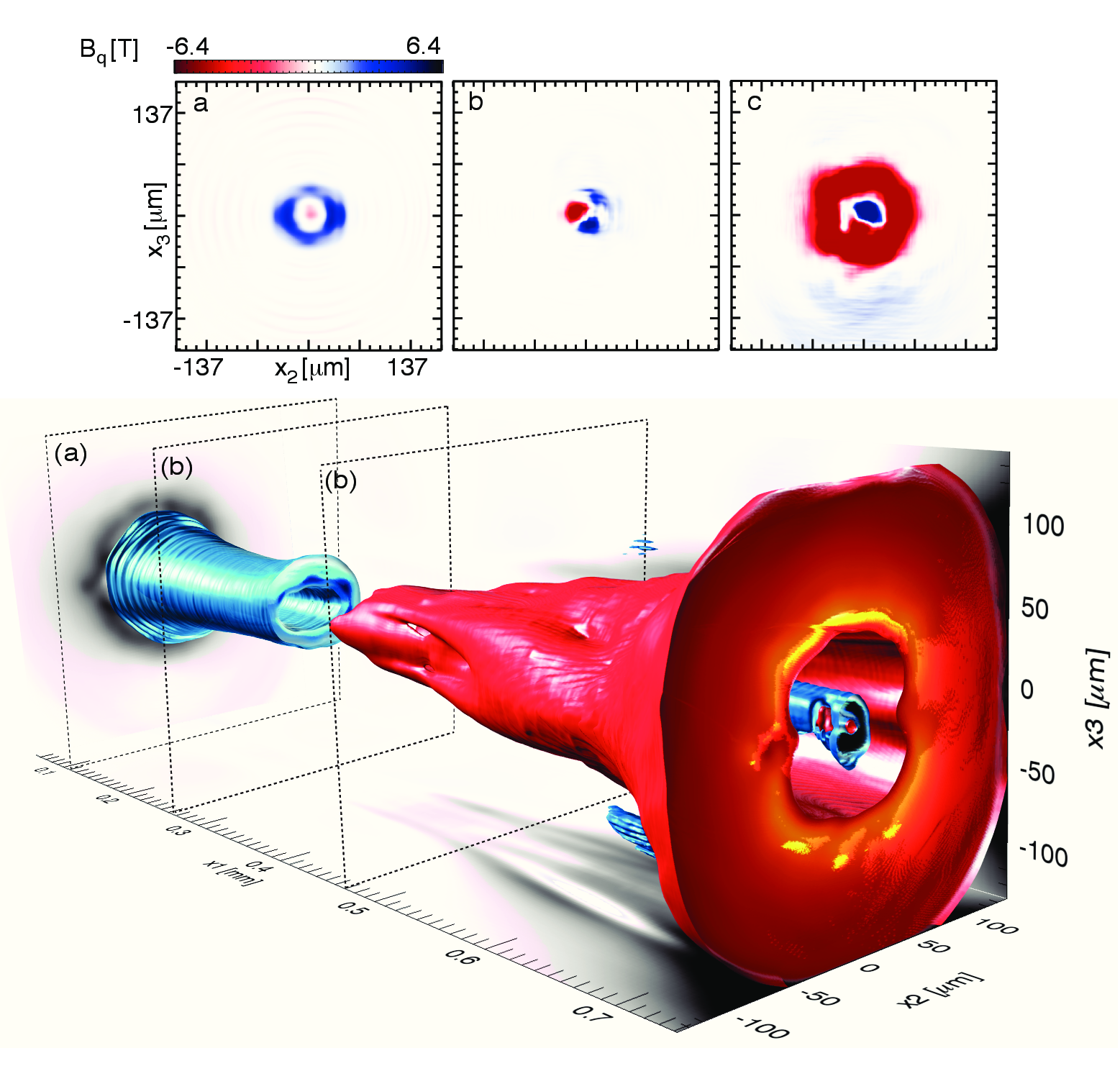}}
			\put(0.5,0.5){(I)}
			\put(5.3,0.5){(II)}
			%\put(0,0){x}
			%\put(10,4.5){x}
		\end{picture}
	\end{center}
	\caption{Summarized results of a full 3D PIC simulation of the experiment (see Methods). The
		final condition of currents (I) and azimuthal magnetic field (II) in the plasma is
		shown after the laser pulse has propagated through the gas; three 2D cuts are shown
	for improved readability.}
	\label{fig:3dsims}
\end{figure*}
%%

%%
%% -------------------------------------------------------------------------------------------

%

An estimate of the poloidal magnetic field amplitude can then be determined considering the return
currents at the plasma-gas interface. Using Ampere's law to estimate the amplitude of the resulting
magnetic field in cylindrical symmetry gives
 $\int \mathbf{B}\cdot\mathrm{d}\mathbf{l} \simeq 2\pi r B_{\theta}  = (4 \pi /c) \int \mathbf{j} \cdot \mathrm{d}\mathbf{S}=\mu_0 \int \mathbf{j}_z \cdot \mathrm{d}\mathbf{S}$, 
where $r$ is the distance to the axis. 
For relativistic hot electrons moving at $c$ longitudinally 
$\mathbf{j}\simeq e\,c\,n_0\,\mathbf{e}_z$, where $e$ is the elementary charge and $\mathbf{e}_z$ is the unit vector pointing in the $z$ direction. 
When $r$ is larger than the plasma radius $r_p$
the return current density flux is $\etahot e\, n_0\, 4\pi^2 \left[r_p^2-(r_p-\Delta)^2\right]\simeq
4 \pi^2 c\, e\, n_0\, \etahot r_p \Delta$, where $\Delta\simeq c/\omega_p \ll r$ is the thickness of
the plasma where the return currents setup, hence $B_{\theta}\simeq 4 \pi e\, n_0\, c\, \etahot
\left(r_p/r\right)\left(c/\omega_p\right)$. 
The typical generated magnetic fields are on the order of $B_{\theta} [T]\simeq 32 \etahot
\sqrt{n_0[10^{16}\si{\cm^{-3}}]}$ for $r\simeq r_p$. Considering $\etahot\simeq 0.05$ (taken from
simulations) and $n_0=\SI{7e-19}{\cm^{-3}}$ we obtain $B_{\theta}\simeq \SI{80}{\tesla}$, which is
consistent with the simulation results. In astrophysical scenarios, the amplitude of the cosmic
magnetic fields generated by this mechanism can be estimated as $0.32 \etahot
\sqrt{n_0[\mathrm{cm^{-3}}]} \,\si{\micro\tesla}$. 

The poloidal magnetic fields observed in our experiment also provide signatures for the occurrence
of wavebreaking, which can be controlled by tuning laser and plasma parameters. The amplitude of the
fields can be controlled by the plasma radius (or the typical size of the ionised cosmic
region) and also by the laser intensity, which determines the fraction of hot electrons (or ultra
high energy particles such as cosmic rays). The orientation of the field indicates the preferential
direction of the hot electron flow.
%

%%
%\setlength{\unitlength}{0.1\linewidth}
%\begin{figure}[!ht]
%\begin{center}
%\includegraphics[width=16cm]{figure5.png}
%\end{center}
%\caption{Main current loops observed during the propagation with the corresponding orientation of the
%	poloidal magnetic field.}
%\label{fig:current_flow}
%\end{figure}
%%%
%
%
%

\begin{small}
\subsection*{Methods}
\begin{description}
	\item[High density micrometric target] 

The target is a pulsed, high pressure, gas jet system which can drive a sub-millimetric gas nozzle
to atomic densities in the range $\num{e19}-\SI{e21}{atoms\per\cm^3}$~\cite{Sylla_RSI11}.

In our experiment we used a transonic nozzle with an output diameter of \SI{400}{\um}, producing an
expanding flow with a Mach number ${\rm M}=1.3$. The radial atomic density
profile at the distance of \SI{200}{\um} from the nozzle level (laser propagation axis) is fit by
$n_A = n_0 \exp\left[-(r/r_0)^{(2.1)}\right]$, where $n_0 = \SI{3.5e19}{atoms\per\cm^3}$ and $r_0 =
\SI{170}{\um}$ is the jet radius. 

The peak density in the jet exponentially decreases along the vertical direction with the
characteristic scale length of \SI{263}{\um}.

	\item[Probing of the plasma] The plasma is probed by a linearly polarized
		$\tau_{\mathrm{p}}=\SI{30}{\fs}$, $\lambda=\SI{400}{\ns}$ probe pulse
		synchronized to the main (pump) beam. The probe beam is split onto three separate
		optical setups, for simultaneous measurement of density and polar magnetic field.

		The radial density map is extracted from the integrated phase of the probe pulse,
		via a Nomarski interferometer on the first transmitted copy of the probe beam.  The
		integrated phase is retrieved by 2D wavelet analysis of the interferogram. The phase
		map is then normalized by subtraction of a reference phasemap (accounting for
		aberrations in the laser transport or in the imaging system) and inverted by
		Hankel-Fourier implementation of the Abel transform.

		Magnetic field is retrieved from the polarization rotation induced on the probe beam
		(Faraday effect) as
		\begin{equation}
			\varphi_{rot} = \frac{e}{2\, m_e\, c\, n_c}\int_{l} 
			n_e\, {\bf B}\cdot{\rm d}{\bf s}
			\label{eq:faraday_int}
		\end{equation}
		where $n_c=m_e \varepsilon_0 \left( 2\pi c / e \lambda \right)$ is the electron
		plasma critical density at the wavelength $\lambda$ (see~\cite{buck_natphys2011}).

		In order to obtain a 2D map of the polarization status in the probe beam profile,
		the pulse is split onto two separate diagnostic lines, each equipped with an analyser and
		a high dynamic range CCD camera. A total of four images, two with pump and two without,
		are used for each snapshot of the magnetic field to eliminate effects due to laser
		intensity fluctuation and systematic optical deformation. The correct superposition of
		images from the two separate polarization measurement lines is ensured by
		spatial markers and automated numerical pattern recognition methods.

		%% - ref. quaderno 2011-I/40
		Considering that the {\it ratio map} $R\left( x, y \right)=I_{p1}/I_{p2}$ can be
		written as
		\begin{equation}
			R\left( x, y \right)\,=\,\frac{1-\beta_1
				\sin^{2}\left( \pi/2 + \theta_{p1} + \varphi_{rot} \right)}{1-\beta_2
					\sin^{2}\left( \pi/2 + \theta_{p2} + \varphi_{rot} \right)},
			\label{eq:def:R}
		\end{equation}
		where $\theta_{p\left\{ 1, 2 \right\}}$ are the analyzers angles, the local
		polarization rotation $\varphi_{rot}\left(x, y \right)$ can be calculated 
		defining $\theta_1 = \pi/2+\theta_{p1}+\varphi_{rot}$ and
		$\Delta_p=\theta_{p2}-\theta_{p1}$ from
		\begin{equation}
			\theta_1\,=\,\tan^{-1}\left[ \frac{a + \left( a^2 + b^2 - c^2
			\right)^{1/2}}{b+c} \right]
			\label{eq:sol:theta1}
		\end{equation}
		where coefficients are given by
		\begin{equation}
			\left\{
			\begin{array}{rcl}
				a & = & R \beta_2\cos\left( 2 \Delta_p \right) - \beta_1\smallskip\\
				b & = & -R \beta_2\sin \left( 2 \Delta_p \right) \smallskip\\
				c & = & R\beta_2 - \beta_1 + 2\left( 1-R \right).
			\end{array}
			\right.
		\end{equation}\\
		The 2D map of the quantity $\left( n_e\,B_{\phi}\right)$ is finally obtained from
		the projected polarization rotation using the appropriate integral
		transform~\cite{flacco_pop2012}:
		\begin{equation}
			\left( n_e B_{\phi} \right)\left( r, x \right)\,=\,\frac{2\,m_e\,c\,n_c}{\pi e}
			\frac{\partial}{\partial r} \int_{r}^{r_0} \frac{\varphi_{rot}\left( y, x
		\right)}{\sqrt{y^2 -r^2}}{\rm d}y.
			\label{eq:frm_inversion}
		\end{equation}
		Due to the sensitivity of the algorithm, $\varphi_{rot}$ maps are
		antisymmetrized before back transformation.

	\item[Simulations] Three-dimensional simulations were performed using the fully relativistic
		partice-in-cell code Osiris~\cite{bib:fonseca_book,bib:fonseca_ppcf_2013}, which is routinely used to model
		laser wakefield acceleration and astrophysical scenarios. The simulation window is $1272\times 300 \times
		300~(\si{\um})^{3}$, divided into $33000\times 1000\times 1000$ cells with $2\times1\times1$
		particles per cell, giving a total of $6.6\times 10^{10}$ simulation particles and a total simulation time between 200000-400000 CPUh. Simulations considered
		an initial Helium gas jet density profile given by
		$n_0=\SI{3.5e19}{\cm^{-3}}\exp\left[\left(\left|x-\SI{700}{\um}\right|/\SI{340}{\um} \right)^{2.1}\right]$, reproducing experimental conditions. Ionisation was modelled
		using ADK tunnel ionisation rates. The longitudinal profile of the laser electric field is symmetric
		and given by $10~\tau^3-15\tau^4+6\tau^5$, with $\tau=\sqrt{2} t/\tau_{\mathrm{FWHM}}$, where
		$\tau_{\mathrm{FWHM}}$ is the FWHM duration of the laser pulse. The transverse laser profile is a
		fit to a transverse line out of the experimental laser intensity profile using higher order
		Hermite-Gaussian modes. Each mode electric field profile is defined as: 
		\begin{equation}
		\begin{array}{rcl}
			E(x) & = & \frac{E_0}{W(x)} H_n\left(\frac{y
					\sqrt{2}}{W(x)}\right)H_m\left(\frac{z\sqrt{2}}{W(x)}\right)\cdot
			\bigskip\\
			&& \cdot\exp\left(\frac{-r^2}{W(x)^2}\right) \cos\left[k_0 x + \frac{k_0 r^2	x}{2(x^2+Z_r^2)} - \zeta_{m,n}(x) \right], 
		\end{array}
		\end{equation} 
		where $E_0$ is the peak electric field, $k_0=\omega_0/c$,
		$\omega_0=\SI{2.34e15}{\radian\per\second}$ is the central laser frequency corresponding to a
		central laser wavelength of $\SI{800}{\nm}$, $\zeta_p(x)=(m+n + 1) \tan^{-1}(x,Z_r)$ is the
		Gouy phase shift, $Z_r=k_0 W_0^2/2$ is the Rayleigh length, $W(x)^2=W_0^2(1+x^2/Z_r^2)$, and $H_n$
		the n-order Hermite polynomial. Table~\ref{tab:laser} shows the laser peak normalised vector
		potentials ($a_0)$ of the five higher order Gaussian beams associated with the fit to the
		experimental laser profile. In addition, $W_0=\SI{9.8}{\um}$ was also used.

		\begin{table}[h!]
		\centering
		\begin{tabular}{lccccc}
			\hline\hline
			(m,n) & $(0,0)$ & $(0,1)$ & $(0,2)$ & $(0,3)$ & $(0,4)$ \\
			\hline
			$a_0$ & 2.27 & -0.076 & -0.05 & $6.3\times10^{-5}$ &  0.025 \\
			\hline\hline
		\end{tabular}
		\caption{\label{tab:laser}Laser intensities for the higher order Hermite-Gaussian modes used to fit the experimental laser profile.}
		\end{table}

\end{description}

\bibliography{miei,sigproc,loaspl,bib,laser-gas,local}
\bibliographystyle{unsrt}

\newpage
\begin{description}
\item[Acknowledgments] The authors acknowledge the support of OSEO project n.I0901001W-SAPHIR, the
support of the European Research Council through the PARIS ERC project (contract 226424) and the
National research grants BLAN08-1-380251 (GOSPEL) and IS04-002-01 (ILA).\\
A.F. acknowledges collaboration from Dr. T. Vinci (LULI, \'Ecole Polytechnique).
Work of JV and LOS partially supported by the European Research Council through the Accelerates ERC
project (contract ERC-2010-AdG-267841) and by FCT, Portugal (contract EXPL/FIZ-PLA/0834/1012). We
acknowledge PRACE for access to resources on SuperMUC (Leibniz Research Center).
\item[Author contributions] S.K., F.S., M.V. and A.F. conceived, designed and carried out the
experimental measurements, A.F.  conceived, designed and realized the analysis tools and performed
the data analysis, A.L., J.V. and L.S. carried out the numerical simulations, A.F., J.V. and L.S.
wrote the manuscript.
\end{description}
\end{small}

\end{document}